\documentclass[twoside,12pt]{article}
\usepackage{amssymb, amsmath, amsfonts}

\newtheorem{lemma}{Lemma}
\newtheorem{definition}{Definition}

\begin{document}

\title{Integrable nonlinear equations on a circle}
\author{ Metin G{\" u}rses$^a$, Ismagil Habibullin\footnote{e-mail:
 habibullin\_i@mail.rb.ru, (On leave from Ufa
Institute of Mathematics, Russian Academy of Science,
Chernyshevskii Str., 112, Ufa, 450077, Russia)}$\,\,^b$
  and Kostyantyn Zheltukhin$^c$ \\
{\small $^a$Department of Mathematics, Faculty of Sciences}\\
{\small Bilkent University, 06800 Ankara, Turkey}\\
{\small e-mail gurses@fen.bilkent.edu.tr}\\
{\small $^b$Department of Mathematics, Faculty of Sciences}\\
{\small Bilkent University, 06800 Ankara, Turkey}\\
{\small e-mail habib@fen.bilkent.edu.tr}\\
{\small $^c$Department of Mathematics, Faculty of Sciences}\\
{\small Middle East Technical University 06531 Ankara, Turkey}\\
{\small e-mail zheltukh@metu.edu.tr}\\}

\begin{titlepage}

\maketitle

\begin{abstract}
The concept of integrable boundary value problems for soliton
equations  on $\mathbb{R}$ and $\mathbb{R}_+$ is extended to 
bounded regions enclosed by smooth
curves. Classes of integrable boundary conditions on a circle for
the Toda lattice and its reductions are found. 
\end{abstract}

{\it PASC: }


\end{titlepage}

\section{Introduction}

The inverse scattering transform method (ISM) discovered in 1967
has proved to be a powerful tool to construct exact solutions and
to solve the Cauchy problem for a large variety of nonlinear
integrable models of mathematical physics. But real physical
applications are usually related to mathematical models with
boundary conditions. For this reason the problem of adopting the
ISM to boundary value problem as well as to initial boundary value
(mixed) problem is very important. During the last two decades
this field of research has been intensively studied. It becomes
clear that only special kinds of boundary conditions preserve the
integrability property of the equation given. Different approaches
were worked out to look for such classes of boundary conditions
based on Hamiltonian structures \cite{s}, on higher symmetries
\cite{ggh}, \cite{ggh1}, \cite{aggh}, and the Lax representation
\cite{ha}, \cite{kg}. Integrable initial boundary value problems
on a half-line (in 1+1 case) or a half-plane (in 1+2 case) for
soliton equations nowadays is a rather studied subject. Analytical
aspects have been developed in \cite{bibt}, \cite{bikt}, \cite{k},
\cite{ahs}, \cite{bi} where large classes of solutions were
constructed. However, boundary value problem for the elliptic
soliton equations or initial boundary value problem for regions
with more complicated boundary are still much less investigated
(see, \cite{bk},\cite{KauJ},\cite{foka1} ).

If the boundary conditions are not consistent with the
integrability property of the equation then the standard version
of the inverse scattering transform method cannot be applied to
the corresponding boundary value problem.  The method requires
very essential modification. Various ideas to extend the ISM to
the initial boundary value problems are suggested in \cite{bk},
\cite{foka3}, \cite{anna2}, \cite{dms}.

In \cite{ha}, \cite{kg} an effective tool to search integrable
boundary conditions has been proposed based on some special
involutions of the auxiliary linear problem. This method (below
for the sake of convenience we refer  it as method of
involutions) can be applied to integrable equations in both 1+1
and 1+2 -dimensional cases. In this article we show that the
method of involutions allows one to extend the concept of
integrability to boundary value problems on bounded regions
enclosed by any closed smooth curve.

Let us explain briefly the approach we use. We call boundary value
problem integrable if it admits a Lax pair. Because of this reason
we look for boundary condition simultaneously with its Lax
representation. The starting point is to make a correct assumption
about the possible form of the Lax pair of the boundary value
problem. Actually this Lax pair is made up from several different
Lax pairs of the original equation itself by gluing the
eigenfunctions along the boundary by properly chosen additional
boundary conditions. As examples  we take Liouville equation and
two-dimensional Toda lattice equation. To generate new Lax pairs
we use point symmetries (involutions) which leave invariant the
nonlinear equation under consideration but change its Lax pair.

In the next  section, as a
trial example, we consider the Liouville equation. We remind the
definition of integrable boundary conditions and find an example
of integrable boundary conditions on a circle. It is shown that
the following nonhomogeneous Neumann problem
\begin{eqnarray}\label{polarLiouville0}
& &u_{rr}+\frac{1}{r} u_r+\frac{1}{r^2}u_{\theta\theta}=8e^u,\\
& &u_r|_{r=a}=-\frac{2}{a}\nonumber
\end{eqnarray}
is integrable, i.e. it  admits a Lax pair (see the list at the end
of the second section).

In the third section we study the two-dimensional Toda lattice
equation on a circular cylinder: $r<a,$ $0\leq\theta\leq2\pi,$
$-\infty<n<\infty$. Several types of integrable boundary value
problems for this lattice are found by using the method of
involutions. Let  $\omega(n)=\exp\{u(n)-u(n+1)\}$. It is shown that the
following boundary values problems are integrable:
\begin{eqnarray}\label{toda01}
&& u_{rr}+\frac{1}{r}u_r+\frac{1}{r^2}u_{\theta\theta}
=\omega(n-1)-\omega(n),\\
&& u(n)|_{r=a}=2in\theta+g(\theta)+k(n),\nonumber
\end{eqnarray}
\begin{eqnarray}\label{toda02}
&& u_{rr}+\frac{1}{r}u_r+\frac{1}{r^2}u_{\theta\theta}
=\omega(n-1)-\omega(n),\\
&& u_r(n)|_{r=a}=\frac{2n}{a}+g(\theta),\nonumber
\end{eqnarray}
\begin{eqnarray}\label{toda03}
&& u_{rr}+\frac{1}{r}u_r+\frac{1}{r^2}u_{\theta\theta}
=\omega(n-1)-\omega(n),\\
&& (u_r(n)+\frac{i}{a}u_\theta(n))|_{r=a}=g(\theta),\nonumber
\end{eqnarray}
It is remarkable that the boundary conditions contain arbitrary
functions $g(\theta)$ and $k(n)$. The Lax pairs for the above
integrable boundary value problems are given in the list at the
end of the third section.

In the fourth section  we consider  periodicity closure
constraints reducing Toda lattice to the  $\sinh$-Gordon and
Tcsitcseika equations. It is shown that the following boundary
values problems are integrable:\\
for the $\sinh$-Gordon equation
\begin{eqnarray}
&& p_{rr}+\frac{1}{r}p_r+\frac{1}{r^2}p_{\theta\theta}=4\sinh p,\\
&& (p_r+\frac{i}{a}p_\theta)|_{r=a}=0; \nonumber
\end{eqnarray}
for the Tcsitcseika equation
\begin{eqnarray}
&& q_{rr}+\frac{1}{r}q_r+\frac{1}{r^2}q_{\theta\theta}=e^{2q}-e^{-q}\\
&& (q_r+\frac{i}{a}q_\theta)|_{r=a}=0. \nonumber
\end{eqnarray}

In the fifth section we give a class of exact solutions of the
Toda lattice on a circle with nonhomogeneous Neumann type boundary
condition on a circle.

\section{Liouville equation}

In this section we concentrate on boundary value problems for
elliptic equations. Suppose that the boundary $\Gamma$ of a domain
$D$ is parameterized by the equation $x'=f(t')$ that introduces a
local system of coordinates by taking $t$-axis along the tangent
direction and $x$-axis along normal to the curve $\Gamma$.

Suppose that the  differential equation under consideration
 \begin{equation}
 E(u)=0
 \end{equation}
admits two different Lax representations. For the sake of
simplicity we take them rewritten in terms of the new coordinates
\begin{equation}\label{L1Liouville}
\begin{array}{l}
Y_x=U(\lambda,u,u_x,\dots)Y(\lambda)\\
Y_t=V(\lambda,u,u_x,\dots)Y(\lambda)\\
\end{array}
\end{equation}
and
\begin{equation}\label{L2Liouville}
\begin{array}{l}
\tilde{Y_x}=\tilde U(\tilde{\lambda} ,u,u_x,\dots)\tilde Y(\tilde\lambda)\\
\tilde{Y_t}=\tilde V(\tilde{\lambda} ,u,u_x,\dots)\tilde Y(\tilde\lambda).\\
\end{array}
\end{equation}
where $\lambda, \tilde \lambda$ are spectral parameters. Now the
equation of the boundary is of the form $x=0$. We are looking for
conditions that allow to relate the equations for $t$ evolution
along the boundary, since $x$ is fixed. More precisely, we have
the following definition
\begin{definition}\label{D1Liouville}
 A boundary condition
\begin{equation}\label{BcondLiouville}
\Omega(t,u,u_t,u_x,\dots)=0
\end{equation}
is integrable if there exists a matrix  $F(\lambda,t,u,...)$ and
function $h(\lambda)$ such that on the boundary $x=0$ the function
$Y=F(\lambda,t,u,...)\tilde Y(\tilde\lambda)$ is a solution of the
equation $Y_t=VY$ for any solution $\tilde Y$ of the equation
$\tilde{Y_t}=\tilde{V}\tilde{Y}$ with $\tilde\lambda=h(\lambda)$,
provided the boundary condition holds.
\end{definition}
If a boundary condition is integrable in the sense of the
definition above this means that the corresponding boundary value
problem admits the Lax representation consisting of the two Lax
pairs (\ref{L1Liouville}) and (\ref{L2Liouville}) defined on the
domain $D$ such that the eigenfunctions $Y$ and $\tilde Y$ satisfy
along the boundary an additional boundary condition $(Y-F\tilde
Y)|_{\Gamma}=0$.

To consider a circle as a boundary we use polar coordinates $(r,
\theta)$. So, the boundary is $r=a$. In polar coordinates the
Liouville equation is
\begin{equation}\label{polarLiouville}
u_{rr}+\frac{1}{r} u_r+\frac{1}{r^2}u_{\theta\theta}=8e^u.
\end{equation}
 It admits the Lax pair
\begin{equation}\label{LaxLiouville}
    Y_r=LY, \quad Y_\theta=AY,
\end{equation}
where $x=r$, $t=\theta$, $U=L$, $V=A$ and
\begin{equation}\label{L-Liouville}
 L=\left(\begin{array}{cc}
   \displaystyle {\frac{e^{u+i\theta}}{2\lambda}+\lambda e^{-i\theta}}&
   \displaystyle { -\frac{e^{u+i\theta}}{2\lambda}+\frac{1}{4}u_r+\frac{i}{4r}u_\theta}\\
   & \\
    \displaystyle { \frac{e^{u+i\theta}}{2\lambda}+\frac{1}{4}u_r+\frac{i}{4r}u_\theta}&
    \displaystyle { -\frac{e^{u+i\theta}}{2\lambda}-\lambda e^{-i\theta}}\\
\end{array}\right)
\end{equation}
\begin{equation}\label{A-Liouville}
 A=ir\left(\begin{array}{cc}
   \displaystyle {\frac{e^{u+i\theta}}{2\lambda}-\lambda e^{-i\theta}}&
   \displaystyle { -\frac{e^{u+i\theta}}{2\lambda}-\frac{1}{4}u_r-\frac{i}{4r}u_\theta}\\
   & \\
    \displaystyle { \frac{e^{u+i\theta}}{2\lambda}-\frac{1}{4}u_r-\frac{i}{4r}u_\theta}&
    \displaystyle { -\frac{e^{u+i\theta}}{2\lambda}+\lambda e^{-i\theta}}\\
\end{array}\right)
\end{equation}
To obtain a second Lax representation we use the Kelvin transformation. The
equation (\ref{polarLiouville}) is invariant under the Kelvin
transformation
\begin{equation}
\bar r=\frac{a^2}{r}\quad \bar u=u+4\ln\frac{a}{r}.
\end{equation}
Under such transformation the Lax pair (\ref{LaxLiouville}) takes
form
\begin{equation}\label{newLaxLiouville}
    \bar Y_r=\bar L\bar Y, \quad \bar Y_\theta=\bar A\bar Y,
\end{equation}
where
\begin{equation}\label{newL-Liouville}
 \bar L=\left(\begin{array}{cc}
   \displaystyle {\frac{r^4e^{u+i\theta}}{2a^4\tilde\lambda}+\tilde\lambda e^{-i\theta}}&
\displaystyle { -\frac{r^4e^{u+i\theta}}{2a^4\lambda}-\frac{r^2u_r}{4a^2}-\frac{r}{a^2}+\frac{ir}{4a^2}u_\theta}\\
 & \\
 \displaystyle { \frac{r^4e^{u+i\theta}}{2a^2\tilde\lambda}-\frac{r^2u_r}{4a^2}-\frac{r}{a^2}+\frac{ir}{4a^2}u_\theta}&
 \displaystyle { -\frac{r^4e^{u+i\theta}}{2a^4\tilde\lambda}-\tilde\lambda e^{-i\theta}}\\
\end{array}\right)
\end{equation}
\begin{equation}\label{newA-Liouville}
\bar A=\frac{ia^2}{r}\left(\begin{array}{cc}
   \displaystyle {\frac{r^4e^{u+i\theta}}{2a^4\tilde\lambda}-\tilde\lambda e^{-i\theta}}&
\displaystyle { -\frac{r^4e^{u+i\theta}}{2a^4\tilde\lambda}+\frac{r^2u_r}{4a^2}+\frac{r}{a^2}-\frac{iru_\theta}{4a^2}}\\
 & \\
    \displaystyle { \frac{r^4e^{u+i\theta}}{2a^4\tilde\lambda}+\frac{r^2u_r}{4a^2}+\frac{r}{a^2}-\frac{iru_\theta}{4a^2}}&
    \displaystyle { -\frac{r^4e^{u+i\theta}}{2a^4\tilde\lambda}+\tilde\lambda e^{-i\theta}}\\
\end{array}\right)
\end{equation}

We will look for a boundary condition under which there exist a
transformation $\tilde \lambda=h(\lambda)$ and non degenerate
matrix $F(\lambda,\theta,u)$ such that $Y(\lambda)=F\tilde
Y(h(\lambda))$ will solve the
$Y_\theta(\lambda)=A|_{r=a}Y(\lambda)$ for every solution $\tilde
Y(\tilde \lambda)$ of the equation $\tilde Y_\theta(\tilde
\lambda)=\tilde A|_{r=a}\tilde Y(\tilde\lambda)$.
\begin{lemma}\label{lemmaLiouville}
The integrable boundary condition  is given by
\begin{equation}\label{Bcond1Liouville}
u_r|_{r=a}=\frac{-2}{a}
\end{equation}
and there are two choices for the matrix $F$ and the function $h$\\
{\it (i)}
\begin{equation}\label{F1Liouville}
h(\lambda)=\lambda, \qquad
 F=\left(\begin{array}{cc}
 1&0\\
 0&1\\
\end{array}\right)
\end{equation}
{\it (ii)}
\begin{equation}\label{F2Liouville}
h(\lambda)=-\lambda, \qquad
 F=\left(\begin{array}{cc}
 0&-1\\
 -1&0\\
\end{array}\right)
\end{equation}
\end{lemma}
{\bf Proof.}   Let $\bar Y(\lambda)$ satisfies the equation $\bar
Y_\theta=\bar A\bar Y$. On the boundary $r=a$, $Y=F\bar
Y(h(\lambda))$ has to satisfy  $Y_\theta= A Y$. Substituting
$Y=F\bar Y(h(\lambda))$ into  $Y_\theta= A Y$ and using $\bar
Y_\theta=\bar A\bar Y$ for $\bar Y_\theta(h(\lambda)$ we obtain
\begin{equation}
\displaystyle{\left(\frac{d}{d\theta}F-A(\lambda) F+F\bar
A(h(\lambda))\right)\bar Y(h(\lambda))=0 }
\end{equation}
The above equality holds if
\begin{equation}\label{conditionLiouville}
\displaystyle{\frac{d}{d\theta}}F=A(\lambda) F-F\bar A(h(\lambda))
\end{equation}
We have an equation for the  unknown  matrix $F$ and function
$h(\lambda)$. To solve the boundary condition
(\ref{BcondLiouville}) with respect to $u_r$ we let
$u_r=G(\theta,u,u_\theta)$. Assuming that $F$ does not depend on
$u_\theta$ and differentiating (\ref{conditionLiouville}) twice with
respect to $u_\theta$ we obtain $\displaystyle{\frac{\partial^2
u_r}{\partial u_\theta^2}}=0$. That is,
$u_r=g_1(u,\theta)u_\theta+g_2(u,\theta)$.
 We substite the above expression for $u_r$ into the
(\ref{conditionLiouville}) and let
\begin{equation}
 F=\left(\begin{array}{cc}
   f_{11}(u,\lambda,\theta)& f_{12}(u,\lambda,\theta)\\
   f_{21}(u,\lambda,\theta)& f_{22}(u,\lambda,\theta)\\
\end{array}\right).
\end{equation} 
 Separating terms with $u_\theta$ and without $u_\theta$ in
(\ref{conditionLiouville}) we obtain two sets of equations. We
write the first set of equations, terms with $u_\theta$, as
\begin{equation}\label{eqnP-Liouville}
\frac{\partial}{\partial u}f=Pf,
\end{equation}
where $f=(f_{11},f_{12},f_{21},f_{22})^T $ and $P$ is a matrix
\begin{equation}\label{P-Liouville}
i a\left(
\begin{array}{cccc}
 0 & - \left(\frac{g_1}{4}-\frac{i}{4 a}\right) &  \left(-\frac{g_1}{4}-\frac{i}{4 a}\right) & 0 \\
 - \left(\frac{g_1}{4}-\frac{i}{4 a}\right) & 0 & 0 &  \left(-\frac{g_1}{4}-\frac{i}{4 a}\right) \\
  \left(-\frac{g_1}{4}-\frac{i}{4 a}\right) & 0 & 0 & - \left(\frac{g_1}{4}-\frac{i}{4 a}\right) \\
 0 &  \left(-\frac{g_1}{4}-\frac{i}{4 a}\right) & -\left(\frac{g_1}{4}-\frac{i}{4 a}\right) & 0
\end{array}
\right)
\end{equation}
We write the second set of equations, terms without $u_\theta$, as
\begin{equation}\label{eqnQ-Liouville}
\frac{\partial}{\partial \theta}f=Qf,
\end{equation}
where $Q$ is a matrix
\begin{equation}\label{Q-Liouville}
ia\left(
\begin{array}{cccc}
\mu-\nu & -\left(\delta+\frac{e^{u+i \theta }}{2
h(\lambda)}\right) &
   \left(-\frac{g_2}{4}-\frac{e^{u+i \theta }}{2 \lambda }\right) & 0 \\
   \\
 - \left(\delta-\frac{e^{u+i \theta }}{2 h(\lambda)}\right) &  \mu+\nu & 0 &
   \left(-\frac{g_2}{4}-\frac{e^{u+i \theta }}{2 \lambda }\right) \\
   \\
  \left(\frac{e^{u+i \theta }}{2 \lambda }-\frac{g_2}{4}\right) & 0 &  -\mu-\nu &
   \left(\delta+\frac{e^{u+i \theta }}{2 h(\lambda)}\right) \\
   \\
 0 &  \left(\frac{e^{u+i \theta }}{2 \lambda }-\frac{g_2}{4}\right) & -\left(\delta-\frac{e^{u+i
  \theta }}{2 h(\lambda)}\right) &  -\mu+\nu\\
\end{array}
\right)
\end{equation}
with $\mu=\displaystyle{\frac{e^{u+i \theta}}{2\lambda }-\lambda
e^{-i \theta } }$,  $\nu=\displaystyle{\frac{e^{u+i \theta }}{2
h(\lambda)}-h(\lambda)e^{-i\theta } }$ and
$\delta=\displaystyle{\frac{g_2}{4}+\frac{1}{a}}$.

 The equations (\ref{eqnP-Liouville}) and (\ref{eqnQ-Liouville}) must be
compatible. This leads to the following compatibility condition
\begin{equation}\label{eqnCompLiouville}
(P_\theta-Q_u+[P,Q])f=0,
\end{equation}
where $[P,Q]$ is a commutator of $P$ and $Q$. The matrix
$(P_\theta-Q_u+[P,Q])$ is nonzero. To have nonzero solution $f$
the determinant of $(P_\theta-Q_u+[P,Q])$ must be zero. It gives
 the following equality
\begin{equation}\label{eqnDetLiouville} \frac{a^6
e^{-4 i \theta }}{16} \left(h^2(\lambda) (a g_1(u,\theta)-i)^2-(a
g_1(u,\theta)+i)^2 \lambda ^2\right)^2=0.
\end{equation}
The above equality holds if either\\
(1) $h(\lambda)=\beta\lambda$ and
$g_1=\frac{i(1+\beta)}{a(1-\beta)}$ where $\beta\in {\mathbb
R}\backslash\{-1,1\}$ or\\
(2) $h(\lambda)=\lambda$ and $g_1=0$ or\\
(3) $h(\lambda)=-\lambda$ and $g_1=0$. \\
One can show that in the case (1) there is no vector $f$ to
satisfy equations (\ref{eqnP-Liouville}) and
(\ref{eqnQ-Liouville}). In the case (2) one has the only solution
$f=q(\lambda)(1,0,0,1)^T$ if $g_2=\frac{-2}{a}$. That gives  the
boundary condition (\ref{Bcond1Liouville}), function $h$ and
matrix $F$ given by (\ref{F1Liouville}).\\
The case (3) is similar to the case (2) and gives same boundary
condition (\ref{Bcond1Liouville}),  function $h$ and matrix $F$
given by (\ref{F2Liouville}).$\Box$

\bigskip

\noindent
 From the above lemma we have the following integrable
boundary value problem with corresponding Lax pairs (we have two
Lax pairs for the problem).

\begin{itemize}
\item $r<a\qquad$ $\displaystyle{u_{rr}+\frac{1}{r}u_r+\frac{1}{r^2}u_{\theta\theta}=8e^u}$,\\
\rule{0pt}{0pt}$\qquad \qquad$ $\begin{array}{l}
 Y_r(\lambda)=L(\lambda)Y(\lambda),\\
 Y_\theta(\lambda)=A(\lambda)Y(\lambda),
 \end{array}$
  and
 $\begin{array}{l}
 \bar Y_r(\lambda)=\bar L(\lambda)\bar Y(\lambda), \\
 \bar Y_\theta(\lambda)=\bar A(\lambda)\bar Y(\lambda),
\end{array}$\\
$r=a\qquad$ $\displaystyle{u_r|_{r=a}=-\frac{2}{a}}$\\
 \rule{0pt}{0pt}$\qquad \qquad$ $Y=F\bar Y$\\
  where $F$ is given by (\ref{F1Liouville}), $L$ is given by (\ref{L-Liouville}), $A$
is given by (\ref{A-Liouville}) and $\bar L$ is given by
(\ref{newL-Liouville}), $\bar A$ is given by
(\ref{newA-Liouville}).
 \item $r<a\qquad$ $\displaystyle{u_{rr}+\frac{1}{r}u_r+\frac{1}{r^2}u_{\theta\theta}=8e^u}$,\\
\rule{0pt}{0pt}$\qquad \qquad$ $\begin{array}{l}
 Y_r(\lambda)=L(\lambda)Y(\lambda),\\
 Y_\theta(\lambda)=A(\lambda)Y(\lambda),
 \end{array}$
  and
 $\begin{array}{l}
 \bar Y_r(-\lambda)=\bar L(-\lambda)\bar Y(-\lambda), \\
 \bar Y_\theta(-\lambda)=\bar A(-\lambda)\bar Y(-\lambda),
\end{array}$\\
$r=a\qquad$ $\displaystyle{u_r|_{r=a}=-\frac{2}{a}}$\\
 \rule{0pt}{0pt}$\qquad \qquad$ $Y(\lambda)=F\bar Y(-\lambda)$\\
  where $F$ is given by (\ref{F2Liouville}), $L$ is given by (\ref{L-Liouville}), $A$
is given by (\ref{A-Liouville}) and $\bar L$ is given by
(\ref{newL-Liouville}), $\bar A$ is given by
(\ref{newA-Liouville}).
\end{itemize}

\section{Two-dimensional Toda Lattice}

We make the same assumption, as in the case of the Liouville
equation, for the coordinates. Hence, boundary is given by $x=0$.
Again we suppose that the differential equation under
consideration admits two different Lax representations
\begin{equation}\label{deflax}
\begin{array}{l}
Y_x=UY\\
Y_t=VY\\
\end{array}
\quad \mbox {and} \quad
\begin{array}{l}
\tilde{Y_x}=\tilde{U}\tilde{Y}\\
\tilde{Y_t}=\tilde{V}\tilde{Y}\\
\end{array}
\end{equation}
For the two dimensional Toda lattice equation 
$U$, $V$, $\tilde U$ and $\tilde V$ in (\ref{deflax}) are linear
operators.
\begin{definition}\label{D1}
 A boundary condition
\begin{equation}\label{Bcond}
\Omega(u)=0
\end{equation}
is integrable if there exists a  linear differential operator $A$
such that on the boundary $x=0$ we have $\tilde Y=AY$ is a
solution of $\tilde{Y_t}=\tilde{V}\tilde{Y}$  for any solution $Y$
of $Y_t=VY$, provided the boundary condition holds.
\end{definition}

To consider a circle as a boundary we use polar coordinates $(r,
\theta)$. So, the boundary is $r=a$. The two dimensional Toda
lattice equation in polar coordinates becomes
\begin{equation}\label{toda}
u_{rr}+\frac{1}{r}u_r+\frac{1}{r^2}u_{\theta\theta}=\omega(n-1)-\omega(n),
\end{equation}
where $\omega(n)=\exp(u(n)-u(n+1))$. The above equation admits a
Lax pair
\begin{multline}\label{laxIU}
\varphi_{1,r}(n)=\displaystyle{\frac{e^{i\theta}}{2}\varphi_1(n+1)-\frac{1}{2}\left(
u_r(n)-\frac{i}{r}u_\theta(n) \right)\varphi_1(n)- }\\
\displaystyle{-\frac{e^{-i\theta}}{2}\omega(n-1)\phi_1(n-1)},
\end{multline}
\begin{multline}\label{laxI}
 \varphi_{1,\theta}(n)=\displaystyle{\frac{ire^{i\theta}}{2}\varphi_1(n+1)-\frac{ir}{2}\left(
 u_r(n)-\frac{i}{r}u_\theta(n)
\right)\varphi_1(n)+}\\
\displaystyle{+\frac{ire^{-i\theta}}{2}\omega(n-1)\varphi_1(n-1)}
\end{multline}
To obtain other Lax representations we use symmetries of the
equation (\ref{toda}).
\begin{enumerate}
 \item Reflection on $\theta$
 \begin{equation}\label{TrTh-Th}
  \tilde{\theta}=-\theta;
 \end{equation}
\item
 The Kelvin transformation
\begin{equation}\label{TrK}
\displaystyle{ \tilde {r}=\frac{a}{r},\quad \tilde {u}=u+4n\ln
\frac{a}{r}};
\end{equation}
\item Reflection on $n$
\begin{equation}\label{Trn-n}
  \tilde {u}=-u(-n).
 \end{equation}
\end{enumerate}
Using the transformation (\ref{TrTh-Th})
we obtain the following Lax representation
\begin{multline}\label{laxIIU}
\varphi_{2,r}(n)=\displaystyle{\frac{e^{-i\theta}}{2}\varphi_2(n+1)-\frac{1}{2}\left(
u_r(n)+\frac{i}{r}u_\theta(n) \right)\varphi_2(n)-}\\
\displaystyle{-\frac{e^{i\theta}}{2}\omega(n-1)\varphi_2(n-1)},
\end{multline}
\begin{multline}\label{laxII}
 \varphi_{2,\theta}(n)=\displaystyle{\frac{ire^{-i\theta}}{2}\varphi_2(n+1)-\frac{ir}{2}\left(
 u_r(n)+\frac{i}{r}u_\theta(n)
\right)\varphi_2(n)+}\\
\displaystyle{+\frac{ire^{i\theta}}{2}\omega(n-1)\varphi_2(n-1)}
\end{multline}
Using the Kelvin transformation (\ref{TrK})
 we obtain the following Lax representation
 \begin{multline}\label{laxIIIU}
\varphi_{3,r}(n)=\displaystyle{\frac{e^{i\theta}}{2}\varphi_3(n+1)-\frac{1}{2}\left(\frac{-r^2}{a^2}u_r(n)+
4n\frac{r}{a^2}-\frac{ir}{a^2} u_\theta
\right)\varphi_4(n)-}\\
\displaystyle{-\frac{r^4e^{-i\theta}}{2a^4}\omega(n-1)\varphi_3(n-1)},
\end{multline}
\begin{multline}\label{laxIII}
\varphi_{3,\theta}(n)=\displaystyle{\frac{ia^2e^{i\theta}}{2r}\varphi_3(n+1)-
\frac{ia^2}{2r}\left(\frac{-r^2}{a^2}u_r(n)+4n\frac{r}{a^2}-\frac{ir}{a^2}u_\theta
\right)\varphi_3(n)+ }\\
\displaystyle{+\frac{ir^3e^{-i\theta}}{2a^2}\omega(n-1)\varphi_3(n-1)}.
\end{multline}
Using the transformations (\ref{Trn-n}) we obtain the following
Lax representation
\begin{multline}\label{laxIVU}
\varphi_{4,r}(n)=\displaystyle{\frac{e^{i\theta}}{2}\varphi_4(n-1)-\frac{1}{2}\left(
-u_r(n)+\frac{i}{r}u_\theta(n) \right)\varphi_4(n)-}\\
\displaystyle{-\frac{e^{-i\theta}}{2}\omega(n)\varphi_4(n+1)},
\end{multline}
\begin{multline}\label{laxIV}
\varphi_{4,\theta}(n)=\displaystyle{\frac{ire^{i\theta}}{2}\varphi_4(n-1)-\frac{ir}{2}\left(-u_r(n)+
\frac{i}{r}u_\theta (n)
\right)\varphi_4(n)+}\\
\displaystyle{+\frac{ire^{-i\theta}}{2}\omega(n)\varphi_4(n+1)}.
\end{multline}
According to Definition 2, to obtain the integrable boundary conditions
we relate the equations for $\theta$
evolution of the above Lax representations, on the boundary $r=a$.
We consider the case when the eigenfunctions are related by multiplication
operator $\varphi_i=A(\theta,n,u,\dots)\cdot\varphi_j$.

It turns out (see the lemma \ref{lemma6}) that that Lax pairs
corresponding to the Kelvin transformation (\ref{TrK}) and the
symmetry (\ref{Trn-n}) are gauge equivalent. A solution of (\ref{laxIII})
transforms to a solution of (\ref{laxIV}) without any boundary
conditions. So, for boundary value problems (\ref{toda02}) and
(\ref{toda03}) we have two possible Lax pairs.

\bigskip

\noindent In the lemma \ref{lemma1} we derive the Lax pair for the
boundary value problem~(\ref{toda01}).
\begin{lemma}\label{lemma1}
Let $\varphi_1(n)$ be a solution of the equation (\ref{laxI}) then
on the boundary $r=a$ a function
$\varphi_2(n)=A\cdot\varphi_1(n)$, where
\begin{equation}\label{AI-II}
A=e^{2in +  g(\theta)} ,\quad g(\theta)\;\mbox {is an arbitrary
function of}\; \theta,
\end{equation}
is a solution of the equation (\ref{laxII}) provided the following
boundary conditions
\begin{equation}\label{boundI-II}
u(n)=2in\theta+g(\theta)+k(n),\quad k(n)\;\mbox {is an arbitrary
function of}\; n,
\end{equation}
holds (for all $n$).
\end{lemma}
{\bf Proof.} On the boundary $r=a$ we substitute
$\varphi_2(n)=A(n,\theta,u,\dots)\cdot\varphi_1(n)$ into the
equation (\ref{laxII}) and use (\ref{laxI}) for
$\varphi_{1,\theta}(n)$. The resulting equation holds if the
coefficients of
 $\varphi_1(n+1)$, $\varphi_1(n)$ and $\varphi_1(n-1)$ are zero. Thus we obtain
the equations
\begin{equation}\label{11}
\displaystyle{\frac{iae^{i\theta}}{2}A(n)=\frac{iae^{-i\theta}}{2}A(n+1)}
\end{equation}
\begin{equation}\label{12}
\displaystyle{A_\theta(n)-\frac{ia}{2}\left(u_r(n)-\frac{iu_\theta(n)}{a}\right)A(n)=
-\frac{ia}{2}\left(u_r(n)+\frac{iu_\theta(n)}{a}\right)A(n)}
\end{equation}
\begin{equation} \label{13}
\displaystyle{\frac{iae^{-i\theta}}{2}\omega(n-1)A(n)=\frac{iae^{i\theta}}{2}\omega(n-1)A(n-1)}
\end{equation}
From the equations (\ref{11}), (\ref{13}) we have that
$A(n)=e^{2i\theta}A(n-1)$. Hence, $A(n)=e^{2i\theta n}b(\theta)$
where $b(\theta)$ is a function of $\theta$ only. Substituting
$A(n)=e^{2i\theta n}b(\theta)$ into the equation (\ref{12}) we
obtain
\begin{equation}
b_\theta+(2in-u_\theta(n))b=0.
\end{equation}
Since the function $b$ does not depend on $n$ we have that the
coefficient of $b$ in the above equation does not depend on $n$,
so $u_\theta(n)=2in +h(\theta)$. Integrating with respect to
$\theta$ we obtain the boundary condition (\ref{boundI-II}), where
$k$ is an arbitrary function of $n$ and $g(\theta)=\int
h(\theta)d\theta$. Then solving the equation (\ref{12}), assuming
that the found boundary condition holds, we obtain $A=e^{2in +
\int g(\theta) d\theta}$, the expression (\ref{AI-II}) for $A$.
$\Box$

\bigskip

\noindent
In a similar way, from the next lemmas we have Lax pairs
for the boundary value problems (\ref{toda02}) and (\ref{toda03}).

\noindent In the lemma \ref{lemma2} and the lemma \ref{lemma3} we
derive the Lax pair for the boundary value problem \ref{toda02}.
\begin{lemma}\label{lemma2}
Let $\varphi_1(n)$ be a solution of the equation (\ref{laxI}) then
on the boundary $r=a$ a function
$\varphi_3(n)=A\cdot\varphi_1(n)$, where
\begin{equation}\label{AI-III}
A=e^{ia\int g(\theta) d\theta},\quad g(\theta)\;\mbox {is an
arbitrary function of}\; \theta,
\end{equation}
is a solution of the equation (\ref{laxIII}) provided the
following boundary conditions
\begin{equation}\label{boundI-III}
u_r(n)=\frac{2n}{a}+g(\theta)
\end{equation}
holds.
\end{lemma}
\noindent {\bf Proof} On the boundary $r=a$ we substitute
$\varphi_3(n)=A(n)\cdot\varphi_1(n)$ into the equation
(\ref{laxIII}) and use (\ref{laxI}) for $\varphi_{1,\theta}(n)$.
The resulting equation holds if the coefficients of
 $\varphi_1(n+1)$, $\varphi_1(n)$ and $\varphi_1(n-1)$ are zero. Thus we obtain
the equations
\begin{equation}\label{21}
\displaystyle{\frac{iae^{i\theta}}{2}A(n)=\frac{iae^{i\theta}}{2}A(n+1)}
\end{equation}
\begin{equation}\label{22}
\displaystyle{A_\theta(n)-\frac{ia}{2}\left(u_r(n)-\frac{iu_\theta(n)}{a}\right)A(n)=
\frac{ia}{2}\left(u_r(n)-\frac{4n}{a}+\frac{iu_\theta(n)}{a}\right)A(n)}
\end{equation}
\begin{equation} \label{23}
\displaystyle{\frac{iae^{-i\theta}}{2}\omega(n-1)A(n)=\frac{iae^{-i\theta}}{2}\omega(n-1)A(n-1)}
\end{equation}
From the equations (\ref{21}), (\ref{23}) we have that $A$ does
not depend on $n$. Hence,  the coefficient of $A$ in (\ref{22})
must be a function of $\theta$ only. This gives the boundary
condition (\ref{boundI-III}). Then solving the equation
(\ref{22}), assuming that (\ref{boundI-III}) holds, we obtain the
expression (\ref{AI-III}) for $A$. $\Box$

\bigskip

\begin{lemma}\label{lemma3}
Let $\varphi_1(n)$ be a solution of the equation (\ref{laxI}) then
on the boundary $r=a$ a function
$\varphi_4(n)=A\cdot\varphi_1(n)$, where
\begin{equation}\label{AI-IV}
A=e^{2i\theta n +u(n)+ia \int g(\theta) d\theta},\quad
g(\theta)\;\mbox {is an arbitrary function of}\; \theta,
\end{equation}
is a solution of the equation (\ref{laxIII}) provided the
following boundary conditions
\begin{equation}\label{boundI-IV}
u_r(n)=\frac{2n}{a}+g(\theta)
\end{equation}
holds.
\end{lemma}
{\bf Proof.} On the boundary $r=a$ we substitute
$\varphi_4(n)=A(n,\theta,u,\dots)\cdot\varphi_1(n)$ into the
equation (\ref{laxIV}) and use (\ref{laxIII}) for
$\varphi_{1,\theta}(n)$. The resulting equation holds if the
coefficients of
 $\varphi_2(n+1)$, $\varphi_2(n)$ and $\varphi_2(n-1)$ are zero. Thus we obtain
the equations
\begin{equation}\label{31}
\displaystyle{\frac{iae^{i\theta}}{2}A(n)=\frac{iae^{-i\theta}}{2}\omega(n)A(n+1)}
\end{equation}
\begin{equation}\label{32}
\displaystyle{A_\theta(n)-\frac{ia}{2}\left(u_r(n)-\frac{iu_\theta(n)}{a}\right)A(n)=
-\frac{ia}{2}\left(-u_r(n)+\frac{iu_\theta(n)}{a}\right)A(n)}
\end{equation}
\begin{equation} \label{33}
\displaystyle{\frac{iae^{-i\theta}}{2}\omega(n-1)A(n)=\frac{iae^{i\theta}}{2}A(n-1)}
\end{equation}
From the equations (\ref{31}), (\ref{33}) we have that
$A(n)=e^{-2i\theta}\omega(n)A(n+1)$. Hence, $A(n)=e^{2i\theta
n+u(n)}b(\theta)$ where $F(\theta)$ is a function of $\theta$
only. Substituting $A(n)=e^{-2i\theta n}b(\theta)$ into the
equation (\ref{32}) we obtain
\begin{equation}
b_\theta-(2in-iau_r(n))b=0.
\end{equation}
Since the function $b$ does not depend on $n$ we have that the
coefficient of $b$ in the above equation does not depend on $n$.
This gives the boundary condition (\ref{boundI-IV}). Then solving
the equation (\ref{32}), assuming that the found boundary
condition holds, we obtain the expression (\ref{AI-IV}) for $A$.
$\Box$

\bigskip

\noindent In the lemma \ref{lemma4} and the lemma \ref{lemma5} we
derive the Lax pair for the boundary value problem \ref{toda03}..
\begin{lemma}\label{lemma4}
Let $\varphi_2(n)$ be a solution of the equation (\ref{laxII})
then on the boundary $r=a$ a function
$\varphi_3(n)=A\cdot\varphi_2(n)$, where
\begin{equation}\label{AII-III}
e^{-2i\theta n + ia\int g(\theta) d\theta},\quad g(\theta)\;\mbox
{is an arbitrary function of}\; \theta,
\end{equation}
is a solution of the equation (\ref{laxIII}) provided the
following boundary conditions
\begin{equation}\label{boundII-III}
u_r(n)=-\frac{i}{a}u_\theta(n)+g(\theta)
\end{equation}
holds.
\end{lemma}
{\bf Proof.} On the boundary $r=a$ we substitute
$\varphi_3(n)=A(n,\theta,u,\dots)\cdot\varphi_2(n)$ into the
equation (\ref{laxIII}) and use (\ref{laxII}) for
$\varphi_{2,\theta}(n)$. The resulting equation holds if the
coefficients of
 $\varphi_2(n+1)$, $\varphi_2(n)$ and $\varphi_2(n-1)$ are zero. Thus we
 obtain the
equations
\begin{equation}\label{41}
\displaystyle{\frac{iae^{-i\theta}}{2}A(n)=\frac{iae^{i\theta}}{2}A(n+1)}
\end{equation}
\begin{equation}\label{42}
\displaystyle{A_\theta(n)-\frac{ia}{2}\left(u_r(n)+\frac{iu_\theta(n)}{a}\right)A(n)=
-\frac{ia}{2}\left(-u_r(n)+\frac{4n}{r}-\frac{iu_\theta(n)}{a}\right)A(n)}
\end{equation}
\begin{equation} \label{43}
\displaystyle{\frac{iae^{i\theta}}{2}A(n)=\frac{iae^{-i\theta}}{2}A(n-1)}
\end{equation}
From the equations (\ref{41}), (\ref{43}) we have that
$A(n)=e^{-2i\theta}A(n-1)$. Hence, $A(n)=e^{-2i\theta n}b(\theta)$
where $b(\theta)$ is a function of $\theta$ only. Substituting
$A(n)=e^{-2i\theta n}b(\theta)$ into the equation (\ref{42}) we
obtain
\begin{equation}
b_\theta-ia(u_r(n)+\frac{i}{a}u_\theta(n))b=0.
\end{equation}
Since the function $b$ does not depend on $n$ we have that the
coefficient of $b$ in the above equation does not depend on $n$.
This give the boundary condition (\ref{boundII-III}). Then solving
the equation (\ref{42}), assuming that the found boundary
condition holds, we obtain the expression (\ref{AII-III}) for $A$.
$\Box$

\bigskip

\begin{lemma}\label{lemma5}
Let $\varphi_2(n)$ be a solution of the equation (\ref{laxII})
then on the boundary $r=a$ a function
$\varphi_4(n)=A\cdot\varphi_2(n)$, where
\begin{equation}\label{AII-IV}
 A=e^{u(n) + \int g(\theta) d\theta},\quad g(\theta)\;\mbox
{is an arbitrary function of}\; \theta,
\end{equation}
is a solution of the equation (\ref{laxIV}) provided the following
boundary conditions
\begin{equation}\label{boundII-IV}
u_r(n)=-\frac{i}{a}u_\theta(n)+g(\theta)
\end{equation}
holds.
\end{lemma}
{\bf Proof.} On the boundary $r=a$ we substitute
$\varphi_4(n)=A(n,\theta,u,\dots)\cdot\varphi_2(n)$ into equation
(\ref{laxIV}) and use (\ref{laxII}) for $\varphi_{2,\theta}(n)$.
The resulting equation holds if the coefficients of
 $\varphi_2(n+1)$, $\varphi_2(n)$ and $\varphi_2(n-1)$ are zero. Thus we obtain  the equations
\begin{equation}\label{51}
\displaystyle{\frac{iae^{-i\theta}}{2}A(n)=\frac{iae^{-i\theta}}{2}\omega(n)A(n+1)}
\end{equation}
\begin{equation}\label{52}
\displaystyle{A_\theta(n)-\frac{ia}{2}\left(u_r(n)+\frac{iu_\theta(n)}{a}\right)A(n)=
-\frac{ia}{2}\left(-u_r(n)+\frac{iu_\theta(n)}{a}\right)A(n)}
\end{equation}
\begin{equation} \label{53}
\displaystyle{\frac{iae^{i\theta}}{2}\omega(n-1)A(n)=\frac{iae^{i\theta}}{2}A(n-1)}
\end{equation}
From the equations (\ref{51}), (\ref{53}) we have that
$A(n)=\omega(n)A(n+1)$. Hence, $A(n)=e^{u(n)}F(\theta)$ where
$F(\theta)$ is a function of $\theta$ only. Substituting
$A(n)=e^{u(n)}b(\theta)$ into the equation (\ref{52}) we obtain
\begin{equation}
b_\theta+(-iau_r(n)+u_\theta(n))b=0.
\end{equation}
Since the function $b$ does not depend on $n$ we have that the
coefficient of $b$ in the above equation does not depend on $n$.
We obtain the boundary condition (\ref{boundII-IV}). Solving the
equation (\ref{52}) and assuming that the found boundary condition
holds, we obtain the expression (\ref{AII-IV}) for $A$. $\Box$

\bigskip

\noindent In Lemma \ref{lemma6} we show that the Lax
representations corresponding to the Kelvin transformation
(\ref{TrK}) and the symmetry (\ref{Trn-n}) are equivalent.
\begin{lemma}\label{lemma6}
Let $\varphi_3(n)$ be a solution of the equation (\ref{laxIII})
then on the boundary $r=a$ a function
$\varphi_4(n)=A\cdot\varphi_2(n)$, where
\begin{equation}\label{AIII-IV}
A=e^{2i\theta n + u(n)}
\end{equation}
is a solution of the equation (\ref{laxIV}).
\end{lemma}
{\bf Proof.} On the boundary $r=a$ we substitute
$\varphi_4(n)=A(n,\theta,u,\dots)\cdot\varphi_3(n)$ into the
equation (\ref{laxIV}) and use (\ref{laxII}) for
$\varphi_{3,\theta}(n)$. The resulting equation holds if the
coefficients of
 $\varphi_3(n+1)$, $\varphi_3(n)$ and $\varphi_3(n-1)$ are zero. Thus we obtain the equations
\begin{equation}\label{61}
\displaystyle{\frac{iae^{i\theta}}{2}A(n)=\frac{iae^{-i\theta}}{2}\omega(n)A(n+1)}
\end{equation}
\begin{equation}\label{62}
\displaystyle{A_\theta(n)-\frac{ia}{2}\left(-u_r(n)+\frac{4n}{a}-\frac{iu_\theta(n)}{a}\right)A(n)=
-\frac{ia}{2}\left(-u_r(n)+\frac{iu_\theta(n)}{a}\right)A(n)}
\end{equation}
\begin{equation} \label{63}
\displaystyle{\frac{iae^{-i\theta}}{2}\omega(n-1)A(n)=\frac{iae^{i\theta}}{2}A(n-1)}
\end{equation}
From the equations (\ref{51}), (\ref{53}) we have that
$A(n)=e^{-2i\theta}\omega(n)A(n+1)$. Hence, $A(n)=e^{2i\theta n
+u(n)}b(\theta)$ where $b(\theta)$ is a function of $\theta$ only.
Substituting $A(n)=e^{2i\theta n+u(n)}b(\theta)$ into the equation
(\ref{52}) we obtain
\begin{equation}
b_\theta=0.
\end{equation}
Hence the function $b$ is a constant. This gives us the expression
(\ref{AII-IV}) for $A$. $\Box$

\bigskip

\noindent From the above lemmas we have  the following list of
integrable boundary value problems with corresponding Lax pairs.
Some of the integrable boundary value problems admit two different
Lax pairs. We give both Lax pairs in the list.
 \begin{center}
 {The list of Lax pairs for two dimensional Toda lattice}
\end{center}
\begin{itemize}
 \item
  $r<a\qquad$ $ \displaystyle{
u_{rr}+\frac{1}{r}u_r+\frac{1}{r^2}u_{\theta\theta}
=\omega(n-1)-\omega(n),\nonumber}$\\
 \rule {0pt}{0pt}$\qquad \qquad \begin{array}{l}
 \varphi_{1,r}(n)=U_1\varphi_1(n),\\
 \varphi_{1,\theta}(n)=V_1 \varphi_1(n),
 \end{array}$
  and
 $\begin{array}{l}
  \varphi_{2,r}(n)= U_2 \varphi_2(n), \\
  \varphi_{2,\theta}(n)= V_2 \varphi_2(n),
\end{array}$ \\
$r=a\qquad$ $ u(n)=2in\theta+g(\theta)+k(n),$\\
    \rule {0pt}{0pt}  $\qquad \qquad \varphi_2=e^{2in +  g(\theta)}\varphi_1 $\\
where action of operator $U_1$ is given by (\ref{laxIU}), $V_1$ is
given by (\ref{laxI}) and $\tilde U_2$ is given by (\ref{laxIIU}),
$\tilde V_2$ is given by (\ref{laxII}).
  \item
 $r<a\qquad$ $\displaystyle{
u_{rr}+\frac{1}{r}u_r+\frac{1}{r^2}u_{\theta\theta}
=\omega(n-1)-\omega(n)}$\\
\rule {0pt}{0pt}
 $\qquad \qquad \begin{array}{l}
 \varphi_{1,r}(n)=U_1\varphi_1(n),\\
 \varphi_{1,\theta}(n)=V_1 \varphi_1(n),
 \end{array}$
  and
 $\begin{array}{l}
  \varphi_{3,r}(n)= U_3 \varphi_3(n), \\
 \varphi_{3,\theta}(n)= V_3  \varphi_3(n),
\end{array}$ \\
$r=a\qquad$ $\displaystyle{u_r(n)=\frac{2n}{a}+g(\theta)}$,\\
\rule {0pt}{0pt} $\qquad\qquad\varphi_3=e^{ia\int g(\theta) d\theta}\varphi_1 $\\
where action of operator $U_1$ is given by (\ref{laxIU}), $V_1$ is
given by (\ref{laxI}) and $U_3$ is given by (\ref{laxIIIU}), $V_3$
is given by (\ref{laxIII}).
 \item
 $r<a\qquad$ $\displaystyle{
u_{rr}+\frac{1}{r}u_r+\frac{1}{r^2}u_{\theta\theta}
=\omega(n-1)-\omega(n)}$\\
\rule {0pt}{0pt}
 $\qquad \qquad \begin{array}{l}
 \varphi_{1,r}(n)=U_1\varphi_1(n),\\
 \varphi_{1,\theta}(n)=V_1 \varphi_1(n),
 \end{array}$
  and
 $\begin{array}{l}
   \varphi_{4,r}(n)= U_4 \varphi_4(n), \\
  \varphi_{4,\theta}(n)= V_4  \varphi_4(n),
\end{array}$ \\
$r=a\qquad$ $\displaystyle{u_r(n)=\frac{2n}{a}+g(\theta)}$,\\
\rule {0pt}{0pt} $\qquad\qquad\varphi_4=e^{2i\theta n +u(n)+ia \int g(\theta) d\theta}\varphi_1 $\\
where action of operator $U_1$ is given by (\ref{laxIU}), $V_1$ is
given by (\ref{laxI}) and
 $U_4$ is given by (\ref{laxIVU}),
$V_4$ is given by (\ref{laxIV}).
 \item
 $r<a\qquad$ $\displaystyle{
u_{rr}+\frac{1}{r}u_r+\frac{1}{r^2}u_{\theta\theta}
=\omega(n-1)-\omega(n)}$\\
\rule {0pt}{0pt}
 $\qquad \qquad \begin{array}{l}
 \varphi_{2,r}(n)=U_2\varphi_2(n),\\
 \varphi_{2,\theta}(n)=V_2 \varphi_2(n),
 \end{array}$
  and
 $\begin{array}{l}
 \varphi_{3,r}(n)= U_3 \varphi_3(n), \\
  \varphi_{3,\theta}(n)= V_3  \varphi_3(n),
\end{array}$ \\
$r=a\qquad$ $\displaystyle{u_r(n)=-\frac{i}{a}u_\theta(n)+g(\theta)}$,\\
\rule {0pt}{0pt} $\qquad\qquad\varphi_3= e^{-2i\theta n + ia\int g(\theta) d\theta}\varphi_2 $\\
where action of operator $U_2$ is given by (\ref{laxIIU}), $V_2$
is given by (\ref{laxII}) and
 $U_3$ is given by (\ref{laxIIIU}),
$V_3$ is given by (\ref{laxIII}).
 \item
 $r<a\qquad$ $\displaystyle{
u_{rr}+\frac{1}{r}u_r+\frac{1}{r^2}u_{\theta\theta}
=\omega(n-1)-\omega(n)}$\\
\rule {0pt}{0pt}
 $\qquad \qquad \begin{array}{l}
 \varphi_{2,r}(n)=U_2\varphi_2(n),\\
 \varphi_{2,\theta}(n)=V_2 \varphi_2(n),
 \end{array}$
  and
 $\begin{array}{l}
 \varphi_{4,r}(n)= U_4 \varphi_4(n), \\
  \varphi_{4,\theta}(n)= V_4  \varphi_4(n),
\end{array}$ \\
$r=a\qquad$ $\displaystyle{u_r(n)=-\frac{i}{a}u_\theta(n)+g(\theta)}$,\\
\rule {0pt}{0pt} $\qquad\qquad\varphi_4= e^{u(n) + \int g(\theta) d\theta}\varphi_2 $\\
where action of operator $U_2$ is given by (\ref{laxIIU}), $V_2$
is given by (\ref{laxII}) and
 $U_4$ is given by (\ref{laxIVU}),
$V_4$ is given by (\ref{laxIV}).
\end{itemize}

\section{Reductions of two dimensional Toda lattice equation}

In this section we  obtain integrable boundary conditions for the
$\sinh$-Gordon and Tcsitcseika equations as reductions of
integrable boundary conditions of the two dimensional Toda
lattice equation.

To reduce the two dimensional Toda lattice equation to the
$\sinh$-Gordon equation we put periodicity condition  $u(n)=u(n+2)$ for all $n$,
where $u$ satisfies (\ref{toda}). Then for $p=u(0)-u(1)$ we have
\begin{equation}
p_{rr}+\frac{1}{r}p_r+\frac{1}{r^2}p_{\theta\theta}=4\sinh p,
\end{equation}
the $\sinh$-Gordon equation in the polar coordinates. Only the
boundary condition of the problem (\ref{toda03}) is consistent with periodicity
constraint $u(n+2)=u(n)$. It gives
\begin{equation}
p_r+\frac{i}{a}p_\theta=0
\end{equation}
on the boundary $r=a$. Evidently by changing $p=iv$ we get
$v_{rr}+\frac{1}{r}v_r+\frac{1}{r^2}v_{\theta\theta}=4\sin v$ and
$v_r+\frac{i}{a}v_\theta=0.$

To reduce the two dimensional Toda lattice equation to the
Tcsitcseika equation we put $u(n)=u(n+3)$ and $u(n)=-u(2-n)$. Then
for $q=u(0)$ we have
\begin{equation}
q_{rr}+\frac{1}{r}q_r+\frac{1}{r^2}q_{\theta\theta}=e^{2q}-e^{-q}
\end{equation}
the  Tcsitcseika equation in polar coordinates. Again only the
boundary condition of the problem (\ref{toda03}) is consistent with periodicity
constraýnt $u(n)=u(n+3)$ and $u(n)=-u(2-n)$ . It gives
\begin{equation}
q_r+\frac{i}{a}q_\theta=0
\end{equation}
on the boundary $r=a$.

\section{Some solutions of the boundary value problem}

\noindent In this section we give an example of solutions for the
special case of the boundary value problem
\begin{equation}\label{bvp1}
u_{rr}+\frac{1}{r^2}u_{\theta\theta}+\frac{1}{r}u_r=\omega(n-1)-\omega(n),\qquad
u_r(n)|_{r=a}=\frac{2n}{a}+g(\theta)
\end{equation}
where $\omega(n)=\exp(u(n)-u(n+1))$. We assume that $g(\theta)=0$ and
look for spherically symmetric solution. That is $u$ is a function
of $r$ only. The boundary value problem (\ref{bvp1}) reduces to
\begin{equation}\label{bvp2}
u_{rr}+\frac{1}{r}u_r=\omega(n-1)-\omega(n),\qquad
u_r(n)|_{r=a}=\frac{2n}{a}.
\end{equation}
Let us introduce new variables  $t=\ln r$ and $v(n,t)=u(n,r)-2n\ln
r$. Then the boundary value problem (\ref{bvp2}) becomes
\begin{equation}\label{bvp3}
v_{tt}=\bar\omega(n-1)-\bar\omega(n),\qquad u_t(n)|_{t=0}=0,
\end{equation}
where $\bar\omega(n)=\exp\{v(n)-v(n+1)\}$. As  solutions of the
above boundary value problem  we can take even solitons of Toda
lattice equation in one dimension. Following \cite{ft}, (see pp.
494-498), the general $N$-soliton solution is given in terms of
the data $\{c, z_j, \gamma_j\}$ such that
\begin{itemize}
 \item[I.] The quantities $z_j$ lie in the interval $-1<z_j<1$ and
are pairwise disjoint.
 \item[II.] $e^{-c}=\prod_{j=1}^N z_j^2$.
 \item[III.] The  quantities $m_j(0)=\displaystyle{\frac{\gamma_j}{\dot{a}(z_j)}}$,
 where $\displaystyle{ a(z)=\prod_{j=1}^N \mbox{sign} z_j  \frac{z-z_j}{zz_j-1}}$
 and dot means derivative with respect to $z$, are positive.
\end{itemize}
The $N$-soliton solution is given by
\begin{equation}\label{soliton}
v(n,t)=c+\ln \frac{\det M(n,t)}{\det M(n-1,t)}
\end{equation}
where $M(n,t)$ is a matrix with entries
$\displaystyle{M_{ij}(n,t)=\delta_{ij}+\frac{\sqrt{m_i(t)m_j(t)}(z_iz_j)^{n+1}}{1-z_iz_j}}$
 and
$m_j(t)=\displaystyle{\frac{e^{-(z_j-z_j^{-1})t}\gamma_j}{\dot a(z_j)}}$, $i,j=1,\dots N$.

The even solitons are describe by the following lemma.
\begin{lemma}
Let $N=2k$ and the data $z_j$, $\gamma_j$, $j=1\dots N$ satisfy
$z_i=-z_{N-i+1}$, $\gamma_i=-\gamma_{N-i+1}$, $i=1\dots k$. Then
the $N$ soliton solution (\ref{soliton}) is even function of $t$.
\end{lemma}
{\bf Proof.} With our choice of the initial data the elements of matrix $M(n,t)$,
that  are
symmetric with respect to the "center" of the matrix, are equal. If
$t$ is changed to $-t$ then every element of  $M(n,t)$  is
replaced by the element symmetric to it with respect the "center"
of the matrix.  Hence determinant of $M(n,t)$ is equal to the
determinant of $M(n,-t)$ and $v(n,t)=v(n,-t)$. From
$v(n,t)=v(n,-t)$ it follows that $v'(t)|_{t=0}=0$.$\Box$

\bigskip

\noindent We give an example of the solutions described in the
above lemma. For $N=2$ we put  $z_1=z_0$, $z_2=-z_0$, $c=-4\ln
z_0$ and $\gamma_1=-\gamma_0$, $\gamma_2=\gamma_0$ where
$0<z_0<1$ and $\gamma_0>0$. Then the data satisfy the conditions
I, II, III and the conditions  of the lemma. With such data one
has the following solution of (\ref{bvp3})
\begin{equation}
v(n,t)=c+\ln\frac{1+\gamma_0(1+z_0^2)z_0^{2n+1}\cosh[
(z_0-z_0^{-1})t] +\gamma_0^2z_0^{4n+4}}
{1+\gamma_0(1+z_0^2)z_0^{2n-1}\cosh[ (z_0-z_0^{-1})t]
+\gamma_0^2z_0^{4n}}
\end{equation}
Hence the boundary value problem (\ref{bvp2}) has the following
solution
\begin{equation}
u(n,r)=c+\ln\frac{1+\frac{1}{2}\gamma_0(1+z_0^2)z_0^{2n+1}(r^{(z_0-z_0^{-1})}+r^{-(z_0-z_0^{-1})})
 +\gamma_0^2z_0^{4n+4}}
{1+\frac{1}{2}\gamma_0(1+z_0^2)z_0^{2n-1}(r^{(z_0-z_0^{-1})}+r^{-(z_0-z_0^{-1})})
+\gamma_0^2z_0^{4n}}
\end{equation}

\section{Conclusion}
In the present paper we apply the method of involutions to
boundary value problems for soliton equations on bounded regions.
As illustrative models we consider Neumann type boundary value
problem on a circle for the Liouville equation and initial
boundary value problem for the two dimensional Toda lattice
equation. The Lax representations for the boundary value problems
are represented.  We considered some reductions of the integrable boundary value
problems in the case of the two dimensional Toda lattice equation.
We also constructed a class of solutions satisfying one of the
found boundary conditions.

\section{Acknowledgments}
The authors thank Scientific and Technical Research Council of Turkey and
Turkish Academy of Science for Partial financial support.

\end{document}